\documentstyle[preprint,aps,epsf,floats,prd]{revtex}

\def\shat{\hat s}
\def\that{\hat t}
\def\uhat{\hat u}
\def\zhat{\hat z}

\begin{document}

\draft

\preprint{\vbox{\hbox{CALT-68-2082}
                \hbox{hep-ph/9610381}
                \hbox{\footnotesize DOE RESEARCH AND}
                \hbox{\footnotesize DEVELOPMENT REPORT} }}

\title{$\psi'$ Polarization due to color-octet quarkonia production}

\author{Adam K. Leibovich}

\address{
Lauritsen Laboratory\\
California Institute of Technology\\
Pasadena, CA 91125}

\date{\today}

\maketitle

\begin{abstract}
We calculated the polarization of $\psi'$ due to $gg \to
Q\bar{Q}[^3P_J^{(8)}]g \to \psi^{(\lambda)}$ color-octet quarkonia
production.  We find that at low transverse momenta the $\psi'$ is
unpolarized due to the contributions proportional to the $L=S=0$ and
$L=S=1$ color-octet matrix elements. As $p_\perp$ increases, the
$\psi'$ mesons become $100\%$ polarized, as predicted by fragmentation
calculations.  Polarization due to lowest order color-singlet
production is also considered, which qualitatively has a similar shape
to the color-octet production.
\end{abstract}


\widetext

The study of quarkonia production has recently received renewed
attention due to new data and interesting theoretical developments.
Traditionally, quarkonia production has been calculated in the
color-singlet model (CSM), where the heavy quark-antiquark pair is
produced in a color-singlet configuration at distance scales short
compared to $\Lambda_{QCD}$.  While the CSM is successful in
describing many phenomenological aspects of quarkonia, it has become
clear that it fails to provide a consistent picture of quarkonia
production.  Order of magnitude discrepancies have been found between
CSM predictions and new measurements of $\psi$ and $\Upsilon$
production at several colliders.  These disagreements have called into
question the validity of the CSM, and stimulated new ideas regarding
quarkonia production.

Quarkonia are inherently nonrelativistic due to the large mass $M_Q$
of the heavy quark and antiquark.  Consequently, the physics of
quarkonia involves a new small parameter, the velocity $v$ of the
heavy constituents inside the $Q\bar{Q}$ bound state.  An effective
field theory called Nonrelativistic Quantum Chromodynamics (NRQCD) has
been established \cite{BBL}, based on a double power series
expansion in the strong interaction fine structure constant $\alpha_s
= g^2_s/4\pi$ and the velocity $v \sim 1/\log M_Q$.  NRQCD allows for
the creation at short distances of a heavy quark-antiquark pair in a
color-octet configuration which later hadronized into a colorless
final state quarkonia.  Similar to Heavy Quark Effective Theory, NRQCD
incorporates an approximate spin symmetry, which constrains various
multiplet structures, transition rates, and polarizations.  There have
been many applications of NRQCD to quarkonia production in various
high energy processes \cite{Braaten}, but the validity of the picture
still has to be verified.

NRQCD makes definite predictions of the polarization of $\psi$'s
produced in a high energy collision.  Thus, one test of the
color-octet picture would be observing the polarization of $\psi$
mesons produced at the Tevatron consistent with NRQCD calculations.
At large transverse momenta, quarkonia are primarily produced by gluon
fragmentation
\cite{BraatenYuan1,BraatenYuan2,BDFM,Roy,Cacciari,BraatenFleming}.
The gluon is nearly real and transverse in the high $p_\perp$ limit,
and the resulting $Q\bar{Q}[^3S_1^{(8)}]$ pair inherits this spin
alignment.  The long distance hadronization into a colorless $\psi$
preserves all angular momentum information, due to the NRQCD
approximate spin symmetry.  Thus, $\psi$ mesons produced at large
$p_\perp$ are 100\% transversely aligned \cite{ChoWise}.  Higher order
$\alpha_s$ correction to the polarization of $\psi$ from gluon
fragmentation have been calculated, and occur at the few percent level
\cite{BenekeRoth1}.

Gluon fragmentation is, however, only valid in the $p_\perp \gg M_Q$
limit.  At low transverse momentum, large numbers of $\psi$'s are
produced via color-octet states with $L=S=0$ and $L=S=1$
\cite{ChoLeibov1,ChoLeibov2}.  Corrections to the fragmentation
limit are not constrained to preserve the polarization of the $\psi$.
Therefore, to use quarkonia polarization as a test of the color-octet
mechanism, we need to investigate the spin alignment due to these
states.  In this paper, we will only consider the polarization of
$\psi'$ mesons.  A similar analysis for other charmonia and bottomonia
states is possible, but these mesons are complicated by feeddown from
higher level states.

In the NRQCD formalism, the production cross section for a quarkonia
state $H$ in the reaction $A+B \to H+X$ can be written as
\begin{equation}\label{crosssection}
d\sigma(A B \to HX) = 
\sum_{ab}\int_0^1dx_1\,dx_2\,f_{a/A}(x_1)\,f_{b/B}(x_2)\,
\sum_n d\hat{\sigma}_{ab}[n] \langle{\mathcal O}^H_n\rangle.
\end{equation}
The first sum in Eq.~(\ref{crosssection}) is over all partons in the
colliding hadrons, and the parton distribution functions are denoted
by $f_{a/A}$ and $f_{b/B}$.  The partonic cross section
$d\hat{\sigma}_{ab}[n]$ describe the production of a quark-antiquark
in a state $n$ and can be calculated perturbatively in $\alpha_s$. The
NRQCD matrix elements $\langle{\mathcal O}^H_n\rangle$ parameterizes
the hadronization of state $n$ into the quarkonium state $H$ plus
light hadrons \cite{BBL}.  These NRQCD matrix elements contain all the
nonperturbative information in the production process, and must be
extracted experimentally.  The order in the velocity expansion at
which each of these matrix elements participates in the $\psi_Q$
creation processes is governed by NRQCD counting rules \cite{Lepage}.

The unpolarized cross sections for producing $\psi'$ mesons in
quark-antiquark, quark-gluon, and gluon-gluon scattering have been
previously calculated up to ${\mathcal O}(\alpha_s^3 v^7)$
\cite{ChoLeibov1,ChoLeibov2}.  At this order, the matrix elements that
appear in Eq.~(\ref{crosssection}) are $\langle {\mathcal O}_1^{\psi'}
(^3S_1)\rangle$, $\langle {\mathcal O}_8^{\psi'} (^3S_1)\rangle$,
$\langle {\mathcal O}_8^{\psi'} (^1S_0)\rangle$, and $\langle
{\mathcal O}_8^{\psi'} (^3P_0)\rangle$.  The values of the color-octet
matrix elements were extracted in \cite{ChoLeibov2,BenekeKramer} by
fitting the magnitudes of the calculated NRQCD cross section to
Tevatron data, and are
\begin{mathletters}\label{matrixelements}
\begin{equation}\label{3s1}
   \langle {\mathcal O}_8^{\psi'} (^3S_1)\rangle = 
   (4.6 \pm 1.0) \times 10^{-3} {\mathrm GeV}^3,
\end{equation}
\begin{equation}\label{1s0and3pj}
   \frac{\langle {\mathcal O}_8^{\psi'} (^3P_0)\rangle}{M_c^2} +
   \frac{\langle {\mathcal O}_8^{\psi'} (^1S_0)\rangle}{3} =
   (5.9 \pm 1.9) \times 10^{-3} {\mathrm GeV}^3.
\end{equation}
\end{mathletters}%
Only the linear combination of $\langle {\mathcal O}_8^{\psi'}
(^1S_0)\rangle$ and $\langle {\mathcal O}_8^{\psi'} (^3P_0)\rangle$
could be extracted.  The error bars are statistical and do not reflect
the large systematic uncertainties in heavy quark masses,
color-singlet wavefunctions, parton distribution functions and
next-to-leading order corrections \cite{BenekeKramer}.  These matrix
elements should only be considered as an order of magnitude estimate,
due to the large uncertainties.  Color-singlet quarkonia production is
suppressed relative to the color-octet production, as can be seen in
Fig.~5 of Ref.~\cite{ChoLeibov2}.

It is possible to use the results of Ref.~\cite{ChoLeibov2} to obtain
the polarization of the final state $\psi'$ when the intermediate
$Q\bar{Q}$ pair is in a ${}^1S_0^{(8)}$ state.  For the intermediate
${}^3S_1^{(8)}$ or ${}^3P_J^{(8)}$ states, however, the analysis in
Ref.~\cite{ChoLeibov2} cannot be used to obtain a polarized cross
section.  As pointed out in Ref.~\cite{BenekeRoth1}, since heavy quark
spin symmetry is an approximate symmetry in the NRQCD Lagrangian,
$L_z$ and $S_z$ are good quantum numbers.  Therefore, to calculate the
polarization of the final state $\psi'$ correctly, we must project the
hard scattering amplitude onto states with definite $L_z$ and $S_z$,
square the amplitude, and then do the sum over $L_z$.  A formalism was
developed later to correctly calculate the polarization in the NRQCD
framework \cite{BraatenChen}.  Also, since we are interested in the
polarization of the quarkonium in the hadron frame, we cannot write
the amplitudes completely in terms of partonic varibles.\footnote{We
thank M. Beneke and M. Kr\"amer for pointing out this error in the
original version of this paper.}  In Ref.~\cite{ChoLeibov2}, the hard
scattering amplitude was projected onto states with definite $JJ_z$
and then squared, with the polarization vectors defined in the parton
frame.  This will give the correct unpolarized cross section
\cite{BenekeRoth2,BraatenChen}, but will not be useful in calculating
the polarized cross section.

The methods used in calculating the amplitudes are similar to those
described in Ref.~\cite{ChoLeibov2}, and the discussion will not be
repeated here.  The only difference being that we projected the
amplitude onto states of definite $L_z$ and $S_z$, squared, and then
summed over $L_z$.  The polarization vector $\epsilon(\lambda)$ of the
quarkonium state then explicitly enters the differential cross section,
Eq.~(\ref{crosssection}), as \cite{BenekeKramer}
\begin{equation}\label{newcrosssection}
\frac{d\hat{\sigma}_{ab}^{(\lambda)}[n]} {d\hat t} = A_{ab}[n] + 
B_{ab}[n] (\epsilon(\lambda)\cdot k_1)^2 + 
C_{ab}[n] (\epsilon(\lambda)\cdot k_2)^2 + 
D_{ab}[n] (\epsilon(\lambda)\cdot k_1)(\epsilon(\lambda)\cdot k_2),
\end{equation}
where $k_1$ and $k_2$ are the momenta of the initial state partons $a$
and $b$.  The coefficients $A,\dots,D$ are shown in the Appendix.  If
we sum over the polarization vector, we recover the unpolarized cross
sections from Ref.~\cite{ChoLeibov2}.

The ratio of longitudinal differential cross section to the
unpolarized differential cross section,
\begin{equation}\label{xidef}
\xi = \frac{\sigma_L}{\sigma_T + \sigma_L},
\end{equation}
can be measured in $\psi'\to \ell^+ \ell^-$ decay.  The leptons are
distributed in angle according to
\begin{equation}\label{angledist}
\frac{d \Gamma (\psi' \to \ell^+ \ell^-)}{d\cos\theta} \propto 
1 + \alpha \cos^2 \theta,
\end{equation}
where
\begin{equation}\label{alphadef}
\alpha = \frac{1-3\xi}{1+\xi},
\end{equation}
and $\theta$ denotes the angle between the lepton momentum in the
$\psi'$ rest frame and the $\psi'$ momentum in the lab frame.  In
Fig.~1, we plot $\alpha$ for prompt $\psi'$ production at the
Tevatron.\footnote{In this paper, MRSD0 parton distribution functions
evaluated at the renormalization scale $\mu = \sqrt{p_\perp^2 +
4M_c^2}$ were used, with $M_c = 1.48$ GeV.  A pseudorapidity cut of
$|\eta| \leq 0.6$ was imposed.}  Since there is only a value for the
linear combination in Eq.~(\ref{1s0and3pj}), we cannot give a definite
prediction for $\alpha$.  Instead, the solid curve represents $\alpha$
when $\langle {\mathcal O}_8^{\psi'} (^3P_0)\rangle = 0$.  The dashed
curve illustrates $\alpha$ when the contribution from $\langle
{\mathcal O}_8^{\psi'} (^1S_0)\rangle$ is set to zero.  The shaded
region illustrates the effect of the uncertainties in the matrix
elements in Eq.~(\ref{matrixelements}).

\begin{figure}[t]
\centerline{\epsfysize=11truecm \epsfbox{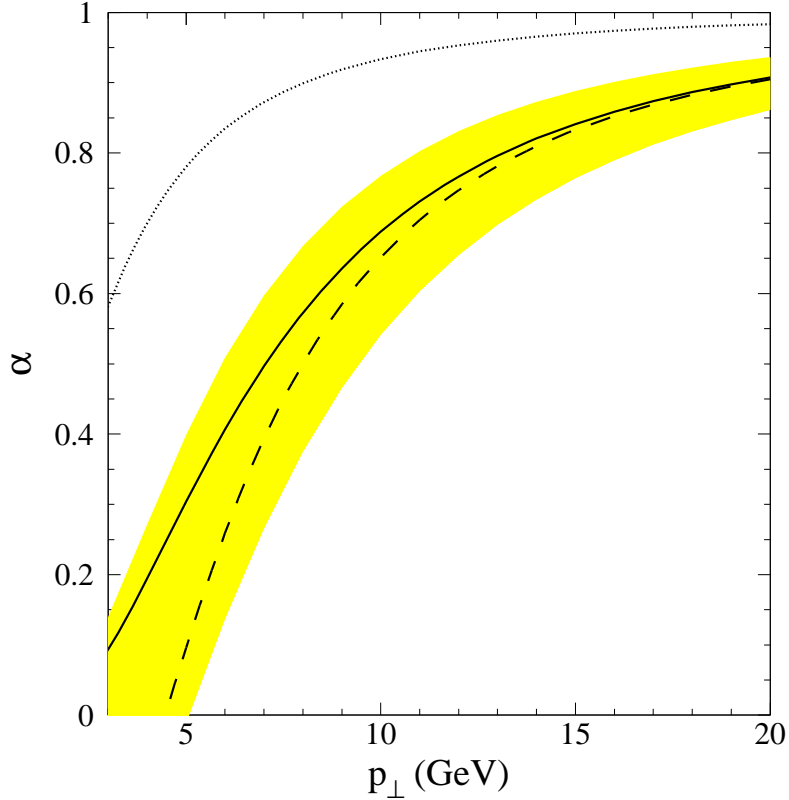}}
\caption[]{Coefficient $\alpha$ which governs the lepton angular 
   distribution in $\psi' \to \ell^+ \ell^-$ decay plotted as a function
   of $p_\perp$.  The solid and dashed curves illustrate $\alpha$ for
   $\psi'$ production at the Tevatron 
   when $\langle {\mathcal O}_8^{\psi'}({}^3P_0) \rangle$ and 
   $\langle {\mathcal O}_8^{\psi'}({}^1S_0) \rangle$
   respectively vanish.  The shaded region shows the effect of the 
   uncertainty in the extraction of the matrix elements.  The dotted 
   line corresponds to the lowest order, color-singlet production.}
\end{figure}

The angular distribution approaches the transverse form $1 + \cos^2
\theta$ at high $p_\perp$ as predicted by gluon fragmentation
computations \cite{ChoWise}. At low transverse momentum, the $\psi'$
is essentially unpolarized due to $L=S=0$ and $L=S=1$ color-octet
states.  Since the two curves are similar in shape, the true value for
the angular coefficient should be close to the curves shown.  The
effect of uncertainties in the matrix elements and higher order
corrections can be qualitatively described by slight displacements of
the curves in Fig.~1, without changing the asymptotic behaviors.
Higher order $\alpha_s$ corrections, however, can change the
asymptotic high $p_\perp$ behavior, but it should be a small effect
\cite{BenekeRoth1}.

Also plotted in Fig.~1 is the polarization due to the lowest order,
color-singlet production, $gg \to c\bar c[\,{}^3S_1^{(1)}] g$.  As can
be seen, the qualitative shape of the color-singlet curve is similar
to the color-octet curves.  While higher order color-singlet
corrections will modify this shape, it is clear that by just observing
the qualitative shape presented in Fig.~1 of the prompt $\psi'$
polarization at the Tevatron will not be a clear signature for
color-octet production.

\acknowledgments

We wish to thank M. Beneke, P. Cho, I.Z. Rothstein, and M. Wise for
useful discussions.  The work of A.K.L. was supported in part by the
U.S. Department of Energy under Grant No. DE-FG03-92-ER40701.

\eject
\appendix
\section*{}

Functions entering into polarized cross sections:\footnote{The
coefficients corresponding to $\langle{\mathcal O}_8^{\psi'}
(^1S_0)\rangle$ production can be obtained from Ref.~\cite{ChoLeibov2}
and will not be repeated here.}

$ q q \to \psi^{(\lambda)} g $:
\begin{mathletters}\label{qq}
\begin{eqnarray}
&&A_{qq}[^3S_1^{(8)}] = 
\frac{4 \alpha_s^3 \pi^2}
{81 M^3 \shat^2}\,
\frac{\left( 4 (\that^2 + \uhat^2) - \that \uhat \right) 
         (\shat^2 - 2 \that \uhat + M^4)}
{\that \uhat (\shat - M^2)^2}, \nonumber\\
&&B_{qq}[^3S_1^{(8)}] = 
-\frac{16 \alpha_s^3 \pi^2}
{81 M \shat^2}\,
\frac{\left( 4 (\that^2 + \uhat^2) - \that \uhat \right)}
{\that \uhat (\shat - M^2)^2}, \\
&&C_{qq}[^3S_1^{(8)}] = 
-\frac{16 \alpha_s^3 \pi^2}
{81 M \shat^2}\,
\frac{\left( 4 (\that^2 + \uhat^2) - \that \uhat \right)}
{\that \uhat (\shat - M^2)^2}, \nonumber\\
&&D_{qq}[^3S_1^{(8)}] = 0,\nonumber \\
&& \nonumber \\
&&A_{qq}[^3P_J^{(8)}] = 
\frac{80 \alpha_s^3 \pi^2}
{27 M^3 \shat^2}\,
\frac{\shat^2 - 2 \that \uhat + 3 M^4}
{\shat (\shat - M^2)^2}, \nonumber \\
&&B_{qq}[^3P_J^{(8)}] = 
-\frac{640 \alpha_s^3 \pi^2}
{27 M \shat^2}\,
\frac{\that \uhat + \uhat M^2 - M^4}
{\shat^2 (\shat - M^2)^3}, \\
&&C_{qq}[^3P_J^{(8)}] = 
-\frac{640 \alpha_s^3 \pi^2}
{27 M \shat^2}\,
\frac{\that \uhat + \that M^2 - M^4}
{\shat^2 (\shat - M^2)^3}, \nonumber \\
&&D_{qq}[^3P_J^{(8)}] = 
\frac{640 \alpha_s^3 \pi^2}
{27 M \shat^2}\,
\frac{\shat^2 + \shat M^2 - 2 \that \uhat}
{\shat^2 (\shat - M^2)^3}, \nonumber
\end{eqnarray}
\end{mathletters}

$ g q \to \psi^{(\lambda)} q $:
\begin{mathletters}\label{gq}
\begin{eqnarray}
&&A_{gq}[^3S_1^{(8)}] = 
-\frac{\alpha_s^3 \pi^2}
{54 M^3 \shat^2}\,
\frac{\left( 4 (\shat^2 + \uhat^2) - \shat \uhat \right) 
         (\that^2 - 2 \shat \uhat + M^4)}
{\shat \uhat (\that - M^2)^2}, \nonumber\\
&&B_{gq}[^3S_1^{(8)}] = 
\frac{2 \alpha_s^3 \pi^2}
{27 M \shat^2}\,
\frac{\left( 4 (\shat^2 + \uhat^2) - \shat \uhat \right)}
{\shat \uhat (\that - M^2)^2}, \\
&&C_{gq}[^3S_1^{(8)}] = 
\frac{4 \alpha_s^3 \pi^2}
{27 M \shat^2}\,
\frac{\left( 4 (\shat^2 + \uhat^2) - \shat \uhat \right)}
{\shat \uhat (\that - M^2)^2}, \nonumber \\
&&D_{gq}[^3S_1^{(8)}] = 
\frac{4 \alpha_s^3 \pi^2}
{27 M \shat^2}\,
\frac{\left( 4 (\shat^2 + \uhat^2) - \shat \uhat \right)}
{\shat \uhat (\that - M^2)^2}, \nonumber \\
&& \nonumber \\
&&A_{gq}[^3P_J^{(8)}] = 
-\frac{10 \alpha_s^3 \pi^2}
{9 M^3 \shat^2}\,
\frac{\that^2 - 2 \shat \uhat + 3 M^4}
{\that (\that - M^2)^2}, \nonumber \\
&&B_{gq}[^3P_J^{(8)}] = 
\frac{80 \alpha_s^3 \pi^2}
{9 M \shat^2}\,
\frac{\shat \uhat + \uhat M^2 - M^4}
{\that^2 (\that - M^2)^3}, \\
&&C_{gq}[^3P_J^{(8)}] = 
\frac{80 \alpha_s^3 \pi^2}
{9 M \shat^2}\,
\frac{\that + M^2}
{\that^2 (\that - M^2)^2}, \nonumber \\
&&D_{gq}[^3P_J^{(8)}] = 
\frac{80 \alpha_s^3 \pi^2}
{9 M \shat^2}\,
\frac{\that^2 - M^2(2 \shat + \that)}
{\that^2 (\that - M^2)^3}, \nonumber
\end{eqnarray}
\end{mathletters}

$ g g \to \psi^{(\lambda)} g $:\footnote{We have introduced the
variable $\zhat = \sqrt{\that \uhat}$ to simplify some of the
coefficients.}
\begin{mathletters}\label{gg}
\begin{eqnarray}
&&A_{gg}[^3S_1^{(1)}] = 
\frac{10 \alpha_s^3 \pi^2 M}
{81 \shat^2}\,
\frac{\shat^2(\shat - M^2)^2 + 
\that \uhat\ (\shat \that + \that \uhat + \uhat \shat - \shat^2)}
{(\shat - M^2)^2 (\that - M^2)^2 (\uhat - M^2)^2} \nonumber \\
&&B_{gg}[^3S_1^{(1)}] = 
-\frac{20 \alpha_s^3 \pi^2 M^3}
{81 \shat^2}\,
\frac{(\shat^2 + \that^2)}
{(\shat - M^2)^2 (\that - M^2)^2 (\uhat - M^2)^2}, \\
&&C_{gg}[^3S_1^{(1)}] = 
-\frac{20\alpha_s^3 \pi^2 M^3}
{81 \shat^2}\,
\frac{(\shat^2 + \uhat^2)}
{(\shat - M^2)^2 (\that - M^2)^2 (\uhat - M^2)^2}, \nonumber \\
&&D_{gg}[^3S_1^{(1)}] = 
-\frac{40\alpha_s^3 \pi^2 M^3}
{81 \shat^2}\,
\frac{\shat^2}
{(\shat - M^2)^2 (\that - M^2)^2 (\uhat - M^2)^2}, \nonumber \\
&& \nonumber \\
&&A_{gg}[^3S_1^{(8)}] = 
\frac{\alpha_s^3 \pi^2}
{36 M^3 \shat^2}\,
\frac{\left[ 27 (\shat^2 - \that \uhat - M^2 \shat) + 19 M^4 \right]}
{(\shat - M^2)^2 (\that - M^2)^2 (\uhat - M^2)^2} \nonumber \\
&& \qquad\qquad\qquad \times
\left[
\shat^2(\shat - M^2)^2 + 
\that \uhat\ (\shat \that + \that \uhat + \uhat \shat - \shat^2) 
\right], \nonumber \\
&&B_{gg}[^3S_1^{(8)}] = 
-\frac{\alpha_s^3 \pi^2}
{18 M \shat^2}\,
\frac{\left[ 27 (\shat^2 - \that \uhat - M^2 \shat) + 19 M^4 \right]
	(\shat^2 + \that^2)}
{(\shat - M^2)^2 (\that - M^2)^2 (\uhat - M^2)^2}, \\
&&C_{gg}[^3S_1^{(8)}] = 
-\frac{\alpha_s^3 \pi^2}
{18 M \shat^2}\,
\frac{\left[ 27 (\shat^2 - \that \uhat - M^2 \shat) + 19 M^4 \right]
	(\shat^2 + \uhat^2)}
{(\shat - M^2)^2 (\that - M^2)^2 (\uhat - M^2)^2}, \nonumber \\
&&D_{gg}[^3S_1^{(8)}] = 
-\frac{\alpha_s^3 \pi^2}
{9 M \shat^2}\,
\frac{\left[ 27 (\shat^2 - \that \uhat - M^2 \shat) + 19 M^4 \right] \shat^2}
{(\shat - M^2)^2 (\that - M^2)^2 (\uhat - M^2)^2}, \nonumber 
\end{eqnarray}
\eject
\begin{eqnarray}
A_{gg}[^3P_J^{(8)}] &=&
\frac{5 \alpha_s^3 \pi^2}
{M^3 \shat^2}
\biggl\{
  M^2 \shat^3 (\shat - M^2)^3  
      (\shat^4  - 2 M^2 \shat^3 + 7 M^4 \shat^2 - 6 M^6 \shat  + 3 M^8)
  \nonumber \\
&& \qquad\quad\,
  + \shat^2 \zhat^2 (\shat - M^2) (\shat^6 - 8 M^2 \shat^5 
     + 23 M^4 \shat^4 - 50 M^6 \shat^3 + 56 M^8 \shat^2 \nonumber \\
&& \qquad\qquad\qquad\qquad\quad\quad
     - 31 M^{10} \shat + 6 M^{12} )
  \nonumber \\
&& \qquad\quad\,
  - \shat \zhat^4 (4 \shat^6  - 9 M^2 \shat^5  + 31 M^4 \shat^4 
    - 71 M^6 \shat^3 + 77 M^8 \shat^2  - 34 M^{10} \shat + 6 M^{12})
  \nonumber \\
&& \qquad\quad\,
  + \zhat^6 (6 \shat^5 + 4 M^2 \shat^4 + 20 M^4 \shat^3  - 33 M^6 \shat^2 
      + 22 M^8 \shat - 3 M^{10})
  \nonumber \\
&& \qquad\quad\,
  - 2 \zhat^8 (2 \shat^3 + 2 M^2 \shat^2 + 5 M^4 \shat - 3 M^6)
  \nonumber \\
&& \qquad\quad\,
  + \zhat^{10} (\shat - M^2)
\biggr \}\bigg/
\left (\shat \zhat^2 (\shat - M^2)^3 (M^2 s + \zhat^2)^3 \right ),
\nonumber \\
%
%
%
B_{gg}[^3P_J^{(8)}] &=&
-\frac{5 \alpha_s^3 \pi^2}
{M \shat^2} \biggl\{
4 \uhat^5 (M^2 - \uhat)^7 \nonumber \\
&&\qquad\qquad\,
- \that \uhat^3 (M^2 - \uhat)^4 
    (M^8 - 7 M^6 \uhat + 42 M^4 \uhat^2 - 52 M^2 \uhat^3 + 24 \uhat^4)
\nonumber \\
&& \qquad\qquad\,
+ \that^2 \uhat^2 (M^2 - \uhat)^3 
    (2 M^{10} - M^8 \uhat - 39 M^6 \uhat^2 + 152 M^4 \uhat^3 
     - 166 M^2 \uhat^4 
\nonumber \\
&& \qquad\qquad\qquad\qquad\qquad\quad\ \,
+ 68 \uhat^5)
\nonumber \\
&& \qquad\qquad\,
- \that^3 \uhat (M^2 - \uhat)^2 
    (M^{12} + 9 M^{10} \uhat + 2 M^8 \uhat^2 - 134 M^6 \uhat^3 
     + 361 M^4 \uhat^4 \nonumber \\
&& \qquad\qquad\qquad\qquad\qquad\quad\ \,
- 339 M^2 \uhat^5 + 116 \uhat^6)
\nonumber \\
&& \qquad\qquad\,
+ \that^4 \uhat (M^2 - \uhat) 
    (11 M^{12} + 9 M^{10} \uhat + 16 M^8 \uhat^2 - 274 M^6 \uhat^3 
     + 589 M^4 \uhat^4 \nonumber \\
&& \qquad\qquad\qquad\qquad\qquad\quad
- 471 M^2 \uhat^5 + 128 \uhat^6)
\nonumber \\
&& \qquad\qquad\,
+ \that^5 (M^2 - \uhat) 
    (4 M^{12} - 51 M^{10} \uhat + 2 M^8 \uhat^2 - 36 M^6 \uhat^3  
     + 282 M^4 \uhat^4 \nonumber \\
&& \qquad\qquad\qquad\qquad\qquad\ \ 
- 329 M^2 \uhat^5 + 80 \uhat^6)
\nonumber \\
&& \qquad\qquad\,
- \that^6 (20 M^{12} - 129 M^{10} \uhat + 94 M^8 \uhat^2 - 12 M^6 \uhat^3 
           + 150 M^4 \uhat^4 - 147 M^2 \uhat^5 
\nonumber \\
&& \qquad\qquad\qquad
+ 8 \uhat^6)
\nonumber \\
&& \qquad\qquad\,
+ 8 \that^7 (5 M^{10} - 19 M^8 \uhat + 6 M^6 \uhat^2 + 6 M^4 \uhat^3 
             - 3 M^2 \uhat^4 + 5 \uhat^5)
\nonumber \\
&& \qquad\qquad\,
- 8 \that^8 
       (5 M^8 - 11 M^6 \uhat - 2 M^4 \uhat^2 + 7 M^2 \uhat^3 - 5 \uhat^4)
\nonumber \\
&& \qquad\qquad\,
+ 20 \that^9 (M^2 - \uhat)^2 (M^2 + \uhat) 
\nonumber \\
&& \qquad\qquad\,
- 4 \that^{10} (M^4 - \uhat^2)
\biggr \}\bigg/
\left ( \shat^2 \that^2 \uhat^2 
(\shat - M^2)^3 (\that - M^2)^3 (\uhat - M^2)^3
\right ),
\nonumber \\
C_{gg}[^3P_J^{(8)}] &=& 
B_{gg}[^3P_J^{(8)}]|_{\that \leftrightarrow \uhat}, \\
D_{gg}[^3P_J^{(8)}] &=& 
\frac{10 \alpha_s^3 \pi^2}
{M \shat^2} \biggl\{
  4 M^2  \shat^6 (\shat - M^2)^5
  \nonumber \\
&& \qquad\quad\ \,
  - M^2 \shat^4 \zhat^2 (\shat - M^2)^2 (22 \shat^3 - 38 M^2 \shat^2 
    + 19 M^4 \shat - 4 M^6) 
  \nonumber \\
&& \qquad\quad\ \,
  - 2 \shat^3 \zhat^4 (\shat^5 - 22 M^2 \shat^4  + 62 M^4 \shat^3 
        - 62 M^6 \shat^2 + 27 M^8 \shat - 4 M^{10})
  \nonumber \\
&& \qquad\quad\ \,
  + \shat^2 \zhat^6 (2 \shat^4 - 17 M^2 \shat^3 + 66 M^4 \shat^2 
     - 31 M^6 \shat + 8 M^8 )
  \nonumber \\
&& \qquad\quad\ \,
  + 2 \shat \zhat^8 (3 \shat^3 - 6 M^2 \shat^2 - 3 M^4 \shat  +  2 M^6)
  \nonumber \\
&& \qquad\quad\ \,
  - 2 \shat \zhat^{10} (5 \shat - 3 M^2) + 4 \zhat^{12}
\biggr \} \bigg/
\left (\shat^2 \zhat^4 (\shat - M^2)^3 (M^2 \shat + \zhat^2)^3
\right ).
\nonumber 
\end{eqnarray}
\end{mathletters}



\begin{references}

\bibitem{BBL} G. T. Bodwin, E. Braaten and G. P. Lepage, Phys. Rev. D 
{\bf 51}, 1125 (1995).

\bibitem{Braaten} See E. Braaten, S. Fleming and T. C. Yuan, 
 Ann. Rev. Nucl. Part. Sci. {\bf 46}, 197 (1996), and references therein.

\bibitem{BraatenYuan1} E. Braaten and T. C. Yuan, Phys. Rev. Lett. {\bf 71}, 
 1673 (1993).

\bibitem{BraatenYuan2} E. Braaten and T. C. Yuan, Phys. Rev. D {\bf 50}, 
 3176 (1994).

\bibitem{BDFM} E. Braaten, M. A. Doncheski, S. Fleming and M. L. Mangano,
 Phys. Lett. {\bf B333}, 548 (1994).

\bibitem{Roy} D. P. Roy and K. Sridhar, Phys. Lett. {\bf B339}, 141 (1994).

\bibitem{Cacciari} M. Cacciari and M. Greco, Phys. Rev. Lett. {\bf 73}, 
 1586 (1994).

\bibitem{BraatenFleming} E. Braaten and S. Fleming, Phys. Rev. Lett. {\bf 74}, 
 3327 (1995).

\bibitem{ChoWise} P. Cho and M. Wise, Phys. Lett. {\bf B346}, 129 (1995).

\bibitem{BenekeRoth1} M. Beneke and I. Z. Rothstein, Phys. Lett. {\bf B372},
 157 (1996).

\bibitem{ChoLeibov1} P. Cho and A. K. Leibovich, Phys. Rev. D {\bf 53}, 
150 (1996).

\bibitem{ChoLeibov2} P. Cho and A. K. Leibovich, Phys. Rev. D {\bf 53}, 
6203 (1996).

\bibitem{Lepage} G. P. Lepage, L. Magnea, C. Nakhleh, U. Magnea and 
K. Hornbostel, Phys. Rev. D {\bf 46}, (1992) 4052.

\bibitem{BenekeKramer} M. Beneke and M. Kr\"amer, Phys. Rev. D {\bf 55},
5269 (1997).

\bibitem{BraatenChen} E. Braaten and Y.-Q. Chen, Phys. Rev. D {\bf 54}, 
 3216 (1996).

\bibitem{BenekeRoth2} M. Beneke and I. Z. Rothstein, Phys. Rev. D {\bf 54},
2005 (1996).

\end{references}
\end{document}